\documentclass[prd,aps,twocolumn,showpacs,preprintnumbers,amsmath,amssymb]{revtex4}
\usepackage[dvips]{graphicx} 
\usepackage{amsmath} 
\usepackage{amssymb} 
 
\usepackage{graphicx}
\usepackage{dcolumn}
\usepackage{bm}
\def\be{\begin{equation}} 
\def\ee{\end{equation}} 
\def\bea{\begin{eqnarray}} 
\def\eea{\end{eqnarray}} 

\newcommand{\comment}[1]{}

\begin{document} 
 
 
\date{\today} 
 
\title{Towards A Nonsingular Tachyonic Big Crunch}

\author{Robert H. Brandenberger \email[email: ]{rhb@hep.physics.mcgill.ca},
Andrew R. Frey \email[email: ]{frey@hep.physics.mcgill.ca} and
Sugumi Kanno \email[email: ]{sugumi@hep.physics.mcgill.ca}}

\affiliation{Department of Physics, McGill University, 
Montr\'eal, QC, H3A 2T8, Canada} 
 
 
\begin{abstract} 

We discuss an effective field theory background
containing the gravitational field, the dilaton and a closed 
string tachyon, and couple this background to a gas of fundamental
strings and D strings. Allowing for the possibility of a
non-vanishing dilaton potential of Casimir type, we demonstrate the 
possibility of obtaining a nonsingular, static tachyon condensate 
phase with fixed dilaton. The time reversal of our solution
provides a candidate effective field theory description of
a Hagedorn phase of string gas cosmology with fixed dilaton.
 
\end{abstract} 
 
\maketitle

\newcommand{\eq}[2]{\begin{equation}\label{#1}{#2}\end{equation}} 
 
\section{Introduction} 

It is widely believed that the ultimate theory of space, time and
matter must resolve the space-time singularities which generically
arise when using Einstein's theory of General Relativity. The
resolution of cosmological singularities is essential in order to
be able to construct a theory of the very early universe which is
genuinely predictive, and the resolution of the singularity inside
of black holes is crucial in order to completely resolve the black hole
information loss problem.

Over the past two years, the challenge of trying to resolve space-like
singularities has been taken up by many researchers working on
string theory. Various new tools have been employed, notably
the AdS/CFT correspondence \cite{AdS,Hertog},
matrix theory \cite{Turok,She}
$c = 1$ matrix theory \cite{Joanna},
matrix string theory \cite{Craps}, 
and tachyon condensation \cite{tachyoncond}.
Whereas the first four approaches are non-perturbative in nature,
the last one is based on perturbative ideas. In this paper, we
will explore the possibility that closed string tachyon condensation 
could lead to a quasi-static, nonsingular final state of cosmology. The
time reversal of this dynamics might lead to a nonsingular
initial state for string cosmology.

Closed string theories often contain states which become tachyonic
when the parameters of the background space on which the perturbative
string theory is set up reach certain values. For example, in
bosonic string theory or in the heterotic string theory (with Scherk-Schwarz
boundary conditions on the fermions),
there are string winding modes which become tachyonic when the size of
a spatial section decreases below some critical value. The 
occurrence of the tachyon signals an instability of the background.
The resulting evolution towards a new ground state can be described
in terms of ``tachyon condensation'' (see e.g. \cite{tachyonrev}
for reviews on closed string tachyon condensation). The tachyonic mode
condenses in a way analogous to how the Higgs field condenses during
the process of spontaneous symmetry breaking. The system will
evolve towards a new and stable ground state which is called the
``tachyon vacuum''.

Closed string tachyon condensation in a cosmological context was
initially considered in \cite{tachyoncond} (see also \cite{Simeon}). It
was argued that fluctuating degrees of freedom disappear once
the tachyon condenses, leading to the effective disappearance
of space itself. It would be nice to be able to understand the
onset of tachyon condensation at the level of an effective
field theory. An initial attempt at analyzing this issue was made in
\cite{Zwiebach}, where an effective field theory containing
gravity, the dilaton and a closed string tachyon was considered. 
In that work, vacuum solutions were considered, and it was
found that, although the string frame metric after tachyon
condensation was static, the dilaton was time-dependent and
tended to strong coupling. 

From early works on string gas cosmology \cite{BV,Perlt}
(see \cite{str,rbr} for recent reviews) it is known that
string matter degrees of freedom, in particular string winding
modes, can play an important role in the cosmological
evolution of the system. In this paper, therefore, we will
couple a gas of stringy degrees of freedom as matter sources
to the background action considered in \cite{Zwiebach} 
and investigate the possibility that the system reaches
a static Hagedorn phase \footnote{The Hagedorn phase is
characterized by a gas of strings in thermal equilibrium close
to the maximal temperature, the Hagedorn temperature \cite{Hagedorn}.}
after the universe contracts. Since we know from \cite{Zwiebach} (in
the context of vacuum solutions) and \cite{Betal} (in the context
of string gas matter) that the dilaton increases towards the strong
coupling regime in the Big Bang/Crunch region (large tachyon and Hagedorn
phases respectively), and thus D-branes
become lighter while the fundamental strings become heavier, we
will include as matter sources both F-strings and D-strings.

We find that in the absence of a dilaton potential, it is
not possible to obtain static solutions. However, if we add
a potential of Casimir type which stabilizes the dilaton, it becomes possible
to find a static solution which represents a candidate for a
nonsingular final state of the universe after contraction.

In the following section, we will derive the cosmological
equations of motion for our dilaton-tachyon background coupled to 
string matter sources. In Section \ref{sec3} we determine some solutions to
these equations. We will show that, provided a suitable potential
for the dilaton is introduced, it is possible to obtain a static
solution with a tachyon condensate. Section \ref{sec4} is devoted to a
discussion of possible implications for cosmology, in particular
for string gas cosmology.

\section{Equations of Motion}\label{s:equations}

Our starting point is the following action $S_g$ for the background 
gravitational field, the dilaton $\Phi$ and the tachyon $T$ (generalized
from the action given in \cite{Zwiebach}):
\bea
S_g \, &=& \, {1 \over {2 \kappa^2}} \int d^4x \sqrt{-g} e^{- 2 \Phi} \\
&& \bigl[R + 4 (\nabla \Phi)^2 - (\nabla T)^2 - V(T) 
- {\tilde V}(\Phi, \lambda) \bigr]
 \, , \nonumber 
\eea
where $g$ is the determinant of the metric, $R$ is the Ricci scalar,
$V(T)$ is the tachyon potential, and we have introduced a dilaton potential
${\tilde V}(\Phi, \lambda)$ which also depends on the scale factor in
a way which could arise if the potential comes from a Casimir-type force. 
We will take the metric to be homogeneous and isotropic, i.e.
\be
ds^2 \, = \, -N^2(t) dt^2 + a(t)^2 \bigl( dx^2 + dy^2 + dz^2 \bigr) \, ,
\ee
where $N(t)$ is the lapse function. Following 
the standard notation \cite{TV}, we will write the
scale factor $a(t)$ in terms of the function $\lambda(t)$ as
\be
a(t) \, = \, e^{\lambda(t)} \, .
\ee
With this ansatz, the background action $S_g$ becomes
\bea
S_g \, &=& \, {1 \over {2 \kappa^2}} \int dt {1 \over N} e^{3 \lambda - 2 \Phi}
\\
&& \bigl[ 3 {\dot \lambda}^2 - ( 3 {\dot \lambda} - 2 {\dot\Phi} )^2 
+ {\dot T}^2 -  N^2 V(T) - N^2 {\tilde V}(\Phi, \lambda) \bigr] \, . \nonumber
\eea

We will couple a gas of string matter to the above background action.
The action for a gas of fundamental strings is
\be
S_{F1} \, = \, - \int dt N F_{F1}(\lambda, \beta N) \, ,
\ee
where $F_{F1}$ is the free energy density of the gas of strings and $\beta$ is
the inverse temperature. Since, in the absence of matter, the dilaton
is dynamical and increases towards the large coupling regime as
tachyon condensation progresses, it is important to include, in addition
to the fundamental strings, a gas of D-strings which become light at
large dilaton values. As can easily be derived from S-duality
considerations, the action $S_{D1}$ of a gas of D-strings is given by
\be
S_{D1} \, = \, - \int dt N e^{ - 2 \Phi} F_{D1}(\lambda - 2 \Phi, \beta N) \, ,
\ee
where $F_{D1}$ is the free energy of the gas of D-strings.

The reader should note that the objects we call D-strings are not the
D1-branes of type II string theory; in particular, for the heterotic
string, stable D-branes do not exist.  However, for compactification on a
6D torus, an NS5-brane wrapped on a
4 dimensions of the torus is S-dual to the F-string
(at tree level), just as
a D-string would be in 10D type IIB string theory \cite{ns5}.  This occurs
because the S-duality used in this paper is S-duality of the 4D dilaton
$\Phi$, which already includes the size moduli of the compactification, 
as opposed to the S-duality of the 10D type IIB string.  For a 
complete thermodynamics invariant under the full $SL(2,Z)$ duality, 
we should also include bound states of the F- and
D-strings; that is, we should consider the gas of all type of string states.

It proves convenient to rewrite the action in terms of the shifted
dilaton $\varphi$ defined via
\be
\varphi \, = \, 2 \Phi - 3 \lambda \, .
\ee
We obtain
\bea
S_g \, &=& \, {1 \over {2 \kappa^2}} \int dt {1 \over N} e^{- \varphi}
\\
&& \bigl[ 3 {\dot \lambda}^2 - {\dot \varphi}^2 
+ {\dot T}^2 - N^2 V(T) - N^2 {\tilde V} \bigr] \, , \nonumber
\eea
where we must keep in mind that ${\tilde V}$ depends on both
$\lambda$ and $\varphi$.

The variational equations of motion with respect to $N$, $\lambda$, $\varphi$
and $T$ then become (after setting $N = 1$):
\bea 
- 3 {\dot \lambda}^2 &+& {\dot \varphi}^2 - {\tilde V} - {\dot T}^2 - V(T) 
\nonumber\\
&=& \, 2 \kappa^2 e^{\varphi} E_{F1} + 2 \kappa^2 e^{-3 \lambda} E_{D1} \, ,
\label{SK}\\
{\ddot \lambda} &-& {\dot \varphi}{\dot \lambda} + 
{ 1 \over 6}{{\partial^{'} {\tilde V}} \over {\partial^{'} \lambda}} 
\nonumber\\
&=& {{\kappa^2} \over 3} e^{\varphi} P_{F1} + \kappa^2 e^{-3 \lambda} F_{D1}
- {{2 \kappa^2} \over 3} e^{-3 \lambda} P_{D1} ,
\label{SF}\\
2 {\ddot \varphi} &-& 3 {\dot \lambda}^2 - {\dot \varphi}^2 - {\dot T}^2 
+ V(T) + {\tilde V} - {{\partial {\tilde V}} \over {\partial \varphi}} 
\nonumber \\
&=& - 2 \kappa^2 e^{- 3 \lambda} F_{D1} 
+ 2 \kappa^2 e^{- 3 \lambda} P_{D1} \, ,
\label{DLT}\\
{\ddot T} &-& {\dot T} {\dot \varphi} 
+{1\over2}{{dV} \over {dT}} \, = \, 0 \, ,
\label{TCN}
\eea
where $E$ and $P$ denote the total energy and pressure, respectively,
and the subscripts indicate whether the terms refer to the contributions
of the fundamental strings or the D-strings, respectively.
Note that in the second equation, the partial derivative $\partial^{'}$
indicates that the derivative is to be taken
at constant value of $\varphi$. Rewritten in terms of partial derivatives
with respect to the original field $\Phi$, we have
\be \label{pdrel}
{{\partial^{'} {\tilde V}} \over {\partial^{'} \lambda}} \, = \,
{{\partial {\tilde V}} \over {\partial \lambda}} + 
{3 \over 2} {{\partial {\tilde V}} \over {\partial \Phi}} \, .
\ee

\section{Solutions of the Equations} \label{sec3}

In the absence of a tachyon and of D-strings, and without a dilaton
potential, our
equations of motion (\ref{SK}-\ref{TCN}) 
reduce to the ones used in the context of
string gas cosmology and which were first discussed in \cite{TV,Ven}:
\bea
- 3 {\dot \lambda}^2 + {\dot \varphi}^2 \, 
&=& \, 2 \kappa^2 e^{\varphi} E_{F1} \, , \label{TV1} \\
{\ddot \lambda} - {\dot \varphi}{\dot \lambda}  
&=& {{\kappa^2} \over 3} e^{\varphi} P_{F1} \, , \label{TV2} \\
2 {\ddot \varphi} - 3 {\dot \lambda}^2 - {\dot \varphi}^2 
&=& \, 0 \, . \label{TV3}
\eea
Adding (\ref{TV1}) to (\ref{TV3}) we obtain 
\be
{\ddot \varphi} - 3 {\dot \lambda}^2 = \, 
\kappa^2 e^{\varphi} E_{F1} \, , \label{TV4} \, .
\ee

From these equations it follows that, if the pressure of the
string gas vanishes, the source term in the equation of
motion (\ref{TV2}) for the scale factor \footnote{Note that we are
working in the string frame.} vanishes. If we consider
the region of phase space in which ${\dot \varphi} < 0$, the
running dilaton acts as friction, and the scale factor approaches
a constant. This situation is realized if we consider a gas of
closed strings on a compact space with stable winding modes \cite{BV,TV}.
In this case (in the context of an expanding universe), 
as we go back in time and the temperature approaches
the Hagedorn temperature, the contributions of the string winding
and string momentum modes combine to give vanishing pressure,
and the string frame scale factor is static. However, from 
(\ref{TV4}) it follows that the dilaton is running, and for the
branch of solutions of interest in string gas cosmology \cite{BV,str,rbr},
the dilaton increases without bound as we go back in time, and thus
rapidly approaches the strong string coupling regime.

In \cite{Zwiebach}, another limiting case of our basic equations
(\ref{SK}-\ref{TCN}) was studied, the limit in which there
is no string matter - only the background fields and the tachyon are
present. In addition, the dilaton potential is set to zero. In
this case, we obtain the following equations:
\bea \label{YZ}
- 3 {\dot \lambda}^2 + {\dot \varphi}^2  - {\dot T}^2 - V(T) \, 
&=& \, 0 \, , \\
{\ddot \lambda} - {\dot \varphi}{\dot \lambda} \, &=& \, 0 \, , \\
2 {\ddot \varphi} - 3 {\dot \lambda}^2 - {\dot \varphi}^2 - {\dot T}^2 
+ V(T) \, &=& 0 \, , \\
{\ddot T} - {\dot T} {\dot \varphi} + {1\over2}{{dV} \over {dT}} \, = \, 0 \, .
\eea

The authors of \cite{Zwiebach} were interested in the dynamics which
occurs once tachyon condensation sets in. Thus, they considered
a tachyon potential $V(T)$ for which Minkowski with fixed dilaton
and vanishing value of the tachyon is an unstable point, i.e.
\be
V(T) \, = \, - {1 \over 2} m^2 T^2 + {\cal O}(T^3) \, .
\ee
It was shown that, during the phase of tachyon condensation,
there are solutions in which the string frame metric is static.
However, as can be seen from the dilaton equation of motion,
the dilaton runs off to the strong coupling regime, and
in fact reaches a strong coupling singularity in finite proper
time. The solutions were interpreted in \cite{Zwiebach} as
corresponding to a big crunch occurring in finite proper time.

If we are interested in the big crunch in cosmology, it is not
justified to neglect matter. If the dilaton has the tendency
to run towards the strong coupling regime, it is also not justified
to focus on fundamental strings as the only matter source.
Hence, we need to add both fundamental strings and their s-duals,
namely D-strings, as matter sources to the system, as we have
done in Section \ref{s:equations}. Since the D-strings have a coupling strength
to the gravitational background which depends inversely on the
dilaton compared to the coupling of F-strings, there is the hope that by
adding both D-strings and F-strings, the dilaton might be
naturally stabilized, even in the absence of a stabilizing
dilaton potential. This idea has already been investigated
by \cite{Subodh}, \cite{Cremonini} and \cite{Rador}. However,
at the level of the effective field theory equations we have
derived (\ref{SK}-\ref{TCN}) this hope is not realized. This can be seen
most easily in the following way. We combine the equation (\ref{SK}) and 
(\ref{DLT}) to obtain (for fixed value of $\lambda$)
\bea \label{nopot}
{\ddot \varphi} - {\dot T}^2 \, &=& \, \kappa^2 e^{\varphi} E_{F1}
+ \kappa^2 e^{- 3 \lambda} E_{D1} \nonumber \\
&& - \kappa^2 e^{- 3 \lambda} F_{D1} + \kappa^2 e^{- 3 \lambda} P_{D1} \, .
\eea
On the other hand, from the equation (\ref{SF}) it follows
that, for vanishing fundamental string pressure, the relation
\be \label{Dpressure}
F_{D1} \, = \, {2 \over 3} P_{D1}
\ee
is required to be consistent with a static solution. Inserting this
relation into (\ref{nopot}) it follows that, for an equation of state
of D-strings with $P > - E$, the right hand side of the equation is
the sum of two positive terms, and thus is inconsistent with the left
hand side being negative. Hence, without the dilaton potential,
the dilaton cannot be time-independent. 

We will now look for solutions to (\ref{SK}) in the presence of 
a non-trivial dilaton potential which have fixed scale
factor and fixed dilaton. This ansatz is consistent provided that
\bea
{\dot T}^2
&=& -V-\tilde V -2\kappa^2 e^{\varphi} E_{F1}\nonumber\\ 
&&- 2 \kappa^2 e^{-3 \lambda} E_{D1}\, ,
\label{SKeq1} \\
{1 \over 6}{{\partial^{'} {\tilde V}} \over {\partial^{'} \lambda}} \,
&=& \, {{\kappa^2} \over 3} e^{\varphi} P_{F1} + \kappa^2 e^{-3 \lambda} F_{D1}
\nonumber \\
&& - {{2 \kappa^2} \over 3} e^{-3 \lambda} P_{D1} \, ,
\label{SKeq2} \\
{\dot T}^2 
&=&V + {\tilde V} - {{\partial {\tilde V}} \over {\partial \varphi}} 
+ 2 \kappa^2 e^{- 3 \lambda} F_{D1} 
\nonumber \\
&& - 2 \kappa^2 e^{- 3 \lambda} P_{D1} \, ,
\label{SKeq3} \\
{\ddot T} + {1\over2}{{dV} \over {dT}} \, &=& \, 0 \, . \label{SKeq4}
\eea

It is straightforward in this case to integrate (\ref{SKeq4}) to find
\be\label{AF1}
\dot T^2 = V_0-V(T)\ ,\ee
where $V_0$ is the value of the potential at the point when $\dot T=0$.
So the tachyon rolling is undamped and unaffected by any other expectation
values in this case.

Adding (\ref{SKeq1}) and (\ref{SKeq3}) yields
\bea \label{SKeq5}
-{{\partial {\tilde V}} \over {\partial \varphi}} \, &=& \,
2\dot T^2 +2 \kappa^2 e^{\varphi} E_{F1} + 2 \kappa^2 e^{-3 \lambda} E_{D1} \\
&& - 2 \kappa^2 e^{- 3 \lambda} F_{D1} + 2 \kappa^2 e^{- 3 \lambda} P_{D1}
\, . \nonumber
\eea
Combining (\ref{SKeq2}) and (\ref{SKeq5}) in order to eliminate the
term containing $F_{D1}$ we obtain
\bea \label{tuning2}
-{{\partial {\tilde V}} \over {\partial \varphi}} + 
{1 \over 3}{{\partial^{'} {\tilde V}} \over {\partial^{'} \lambda}}&=& 
2\dot T^2 +2 \kappa^2 e^{\varphi} E_{F1}
+ 2 \kappa^2 e^{-3 \lambda} E_{D1}\nonumber\\
&& + {{2 \kappa^2} \over 3} e^{\varphi} P_{F1} 
+ {{2 \kappa^2} \over 3} e^{-3 \lambda} P_{D1} \, .
\eea
Making use of (\ref{pdrel}) to express the partial derivative for fixed
$\varphi$ in terms of partial derivatives with fixed $\Phi$, we find that
only the contribution of the second term in (\ref{pdrel}) survives, and 
the left-hand side of (\ref{tuning2}) becomes
\be
{1 \over 3} {{\partial {\tilde V}} \over {\partial \lambda}} \, .
\ee
We thus see that a dilaton potential which is independent of $\lambda$ is
not sufficient to stabilize the dilaton. A Casimir-type potential
is required in order to obtain a Hagedorn phase with fixed dilaton. A
potential of the form
\be
{\tilde V}(\Phi, \lambda) \, = \, e^{-\gamma \lambda} W(\Phi) \,
\ee
or 
\be
\tilde V (\Phi,\lambda)\, =\, W e^{-\gamma\lambda} +\tilde W(\Phi) \ee
would have the right sign of its partial derivative (if $W$ is negative)
to make it
possible to satisfy (\ref{tuning2}) without any violation of the
usual energy conditions.

Next, from (\ref{SKeq1}) we see that, provided that the 
potential adjusts itself to cancel the energy densities of the matter
gases and the tachyon velocity, the Hamiltonian constraint equation
is satisfied:
\be \label{tuning1}
-V -\tilde V\, = \, {\dot T}^2 + 2 \kappa^2 e^{\varphi} E_{F1} 
+ 2 \kappa^2 e^{ - 3 \lambda} E_{D1} \, .
\ee 
This of course requires a negative cosmological constant.

An important issue is the stability of the solution we have found.
We have not yet performed a complete stability analysis. An initial
investigation shows that, in the limit of negligible tachyon mass,
the solution is stable towards linear
perturbations provided that the dilaton potential obeys a further
constraint on the difference of its first and second derivatives.
We hope to return to this issue in the future.
 
In conclusion, we have shown that, provided the tunings (\ref{tuning2})
and (\ref{tuning1}) for the tachyon and dilaton potentials are
satisfied, then it is possible to obtain a tachyon condensate background
in which both the scale factor and the dilaton are static. 
The tuning (\ref{tuning1}) of the tachyon potential depends on
the energy of the system. We can
try to interpret this tuning as an adjustment of the
cosmological constant. However, it does not appear natural at the level
of the current effective field theory analysis that this tuning
should depend on the matter content of the universe. However, if
the effective cosmological constant adjusts itself dynamically
as in the scenarios of \cite{RHBrev5} and \cite{Stephon}, then one
might imagine that such an adjustment might be possible.

\section{Connection to Cosmology} \label{sec4}

The picture of tachyon condensation in a collapsing universe is as
follows: initially, the universe is decreasing in size and no tachyon
is present. Once the size has reached some critical value, the
tachyonic modes begin to condense, and the field $T$ begins to roll
away from $T = 0$, the local maximum of its effective potential. In
the context of our effective action, we expect that a nonsingular 
tachyon condensate
phase should have constant scale factor and dilaton.  We have only been able
to find this behavior by introducing a dilaton potential with
additional Casimir-like dependence on the scale factor.

The time reversal of this nonsingular tachyon big crunch 
would be a candidate for a
nonsingular Hagedorn phase with fixed dilaton. In the context
of this background, one would then be able to realize the
recently suggested \cite{NBV,BNPV2} ``string gas cosmology
structure formation scenario" in which thermal fluctuations
of a gas of closed strings during the Hagedorn phase generate
a nearly scale-invariant spectrum of adiabatic cosmological 
fluctuations with a slight red tilt and a spectrum of gravitational
waves with a slight blue tilt \cite{BNPV1}. In the following,
we first present a brief review of the new structure formation
scenario \footnote{Note that this scenario provides an alternative
to inflation in explaining the currently observed cosmological
data on matter inhomogeneities and cosmic microwave anisotropies.}.
In particular, we comment on why our new background satisfies 
the criteria required to make the scenario of \cite{NBV} work.

String gas cosmology \cite{BV,Perlt} is an attempt to
construct a toy model of the early universe incorporating key
symmetries and new degrees of freedom which distinguish string
theory from point particle field theory. The key symmetry
which has been made use of extensively in string gas cosmology
is T-duality, the key new stringy degrees of freedom are
string oscillatory modes and closed string winding modes,
neither of which are present in field theories. The presence
of string oscillatory modes leads to the existence of a 
limiting temperature, the Hagedorn temperature \cite{Hagedorn}
for a gas of strings in thermal equilibrium. 

In the following, we will imagine that all spatial sections are
compact, admitting stable one cycles. In this case, string
modes which wind the spatial cycles are stable. String states
are thus characterized by their center of motion momentum
quantum number, an integer $n$, the winding number $m$ (another
integer), and the quantum numbers describing string oscillations.
The T-duality symmetry is the invariance of the spectrum of
string states under the interchange $R \rightarrow 1/R$, where
$R$ is the radius of the spatial sections (in units of the
string length), provided that the quantum numbers $n$ and $m$ are
also interchanged. Based on this symmetry, a universe with
radius $R$ is equivalent, at least in terms of string thermodynamics,
to a universe of radius $1/R$ \cite{BV}. 

Based on the above string thermodynamic analysis, it thus appears
reasonable to expect that string theory should lead to a nonsingular
cosmology, where $R = 1$ is a pivot point of the dynamics between
a branch of cosmological expansion and another branch of cosmological
contraction which, in the correct units, to an observer looks
identical to that of an expanding universe \cite{BV}. The state
of the system at this pivot point should correspond to a dense
gas of strings at a temperature close to the Hagedorn temperature.
We call this the ``Hagedorn phase".

It is obvious that the correct description of the Hagedorn phase
will require the knowledge of non-perturbative string theory.
In the absence of such a theory, we are forced to develop toy
models of a classical background space-time coupled to a gas
of strings. This is what we mean by ``string gas cosmology".

It is clear from the outset that our background cannot be
described by General Relativity (GR) since GR violates the
T-duality symmetry. A key requirement for a string gas
cosmology background is that it be able to describe the
transition from a static Hagedorn phase to the expanding
radiation phase of standard cosmology without requiring
matter which violates the weak energy condition \footnote{Thus
we disagree with the criticisms expressed
in \cite{KKLM}.}. In a lot of the work on string gas
cosmology (beginning with \cite{TV}), 
dilaton gravity was used as a better justified
background since it includes the dilaton at the same level
as the graviton, as should be the case in a string theory
model. 

As we have already seen in Section \ref{sec3}, in dilaton gravity,
whereas the string frame scale factor is indeed constant
in the Hagedorn phase, the dilaton is not constant. In
fact, following the radiation phase of standard cosmology
into the past in the context of the dilaton gravity equations,
we find that the dilaton begins to diverge towards strong
coupling when the string gas approaches the Hagedorn
temperature. Thus, in dilaton gravity the expected
symmetry between large $R$ and small $R$ is not realized.
Another way to put this problem is the following: the dilaton
gravity description of the Hagedorn phase is in conflict
with another key symmetry of string theory, namely S-duality.

A background in which both the dilaton and the scale factor
are constant in the Hagedorn phase (and, in consequence,
the string frame and the Einstein frame coincide),
will yield a better description of what we believe the
true stringy description of the Hagedorn phase will look like.
Therefore, we have been motivated to include a potential that stabilizes
the dilaton near the S-duality fixed point.

A background with constant dilaton is also crucial in order
to implement the string gas structure formation scenario of
\cite{NBV}. In this scenario, thermal fluctuations of a
gas of closed strings induces the cosmological perturbations.
During the Hagedorn phase, the Hubble radius is infinite,
and thus there are no obstacles to the gradual development
of thermal equilibrium. At the end of the Hagedorn phase
string winding modes decay into radiation and the pressure
of the string gas increases from zero to the value it has in
the radiation phase. Thus, at the end of the Hagedorn phase,
the Hubble radius rapidly decreases to a microscopic value,
and fixed comoving scales quickly exit the Hubble radius. Matter
fluctuations established in the Hagedorn phase induce
metric fluctuations (see \cite{MFB} for a comprehensive
review of the theory of cosmological perturbations, and
\cite{RHBrev2} for a pedagogical summary).
Provided that the gas of strings contains stable winding
modes, the specific heat of a region scales as $R^2$ with the radius $R$ of
the region~\footnote{The precise thermodynamical calculation confirming
this behavior awaits future work.}. 
If the metric fluctuations can be inferred from
the matter fluctuations at Hubble radius crossing making
use of the Einstein constraint equations, then a roughly
scale-invariant spectrum of cosmological perturbations with
a small red tilt results. However, as discussed in detail
in \cite{Betal}, in dilaton gravity the dilaton fluctuations
cannot be neglected and lead to a blue spectrum (see also
\cite{KKLM}). This emphasizes the need to stabilize the dilaton.

In Section \ref{sec3} we have constructed a background which indeed
features both constant scale factor and dilaton, and which
thus is a candidate for an improved description of the
Hagedorn phase. A key question to ask, therefore, in order
to make contact with the work of \cite{NBV,BNPV2} is
how the dynamics described in Section \ref{sec3} can be realized
in the context of an expanding cosmology. 

We imagine starting the universe in the tachyon condensate
phase discussed in this paper, but with a tachyon which is
climbing up its potential instead of rolling down as it does
when we are considering a collapsing universe. One can then
imagine that when the tachyon reaches the local maximum of
its potential at $T = 0$, the tachyon disappears. If the dilaton
is fixed by its potential, we will then be emerging into an
expanding universe. In this context, the tachyon phase can
provide the {\it strong coupling Hagedorn phase} postulated
in \cite{Betal}.

\section{Conclusions}
 
In this paper we have constructed an effective field theory
background which leads to a nonsingular big crunch. Our
background includes usual gravity, the dilaton, and a tachyon
condensate. Matter is described by a gas of closed fundamental
strings and also includes a gas of D-strings. We have shown
that, provided we introduce a non-vanishing dilaton potential
with Casimir-type dependence on the volume,
and tune both the tachyon and dilaton potential to appropriate
values, a phase with constant scale factor and constant
dilaton is possible. (The presence of the tachyon itself neither
helps nor hinders the stabilization of the dilaton.)
Such a state can be obtained via
tachyon condensation from a contracting phase of standard
cosmology.  

Our construction may appear rather baroque. However, we should
keep in mind that, if we start with a gravitational
action which is based on Einstein or dilaton gravity, we are
starting from a point far removed from one in which all of
the symmetries expected from string theory are manifest. 
Our work is therefore intended more as an ``existence proof"
which might shed some light on what the key ingredients of
the ultimate description of the Hagedorn phase might be
\footnote{See \cite{Bogo} for an interesting attempt
at constructing a background for string gas cosmology making
use of a theory with two metrics in which the symmetries of
string theory are more manifest.}.

We have speculated that the time reversal of our big crunch
construction could provide a good description of an initial
Hagedorn phase of string gas cosmology with fixed scale
factor and dilaton in which the recently proposed string
gas structure formation scenario \cite{NBV,BNPV2} can be realized.
A concern at this point is that in \cite{NBV,BNPV2} it is
assumed that the cosmological fluctuations can be described using
the Einstein equations. What ensures that this is the case
given our modified dynamics?

A concrete toy model in which this question can be analyzed is
the higher derivative, ghost-free and asymptotically free
higher derivative gravity model of \cite{Tirtho1}. 
Cosmological solutions of this theory lead to a bouncing cosmology.
The dilaton is fixed in this model. The dynamics away from the
bounce point rapidly approaches that of standard cosmology. As
discussed in \cite{Tirtho2}, this model provides a background in
which the string gas structure formation scenario can be realized.
It is justified to use the equations of motion of Einstein gravity
to describe the evolution of cosmological perturbations near the
bounce because the effects of the higher derivative terms in
the action are suppressed by powers of $k/M$, where $k$ is
the physical momentum scale of the fluctuation mode, and $M$ is
the cutoff scale. The length scale corresponding to $k$ is, for
scales of interest in cosmology today, about $1 {\rm mm}$,
whereas the length scale corresponding to $M$ will be given by
the string length. Thus, the correction terms will have a negligible
effect.

Returning to our scenario, it the dilaton remains frozen during the
transition between the tachyon condensate phase and the radiation phase
of standard cosmology, a transition induced by a change in the equation of
state of the strings (for example triggered by the decay of string 
winding modes into string loops), then the scale factor will begin to
increase, thus providing the cosmological background required for
the structure formation scenario of \cite{NBV}. Since the dilaton is
fixed, it is justified to use the usual equations of motion of
the theory of cosmological perturbations in the Einstein frame to
follow the evolution of the fluctuations.

\begin{acknowledgments} 
 
We wish to thank Simeon Hellerman, Maulik Parikh,
Joe Polchinski and Jiro Soda 
for fruitful discussions. One of us (RB) acknowledges
hospitality and financial support by the KITP in Santa Barbara during
the mini-program on the {\it Quantum Nature of Spacetime Singularities}
in January 2007 during which some of the work on this project was done.
AF is supported in part by a Institute for Particle Physics postdoctoral
fellowship, and SK is supported by a grant for research abroad by the
JSPS. The work of RB and AF is supported in part by an 
NSERC Discovery Grant and by the Canada Research Chairs Program. RB is
also supported in part by funds from a FQRNT Team Grant.

\end{acknowledgments}


\begin{thebibliography}{99} 
 
\bibitem{AdS}
S.~R.~Das, J.~Michelson, K.~Narayan and S.~P.~Trivedi,
  ``Cosmologies with null singularities and their gauge theory duals,''
  Phys.\ Rev.\  D {\bf 75}, 026002 (2007)
  [arXiv:hep-th/0610053];\\
S.~R.~Das, J.~Michelson, K.~Narayan and S.~P.~Trivedi,
  ``Time dependent cosmologies and their duals,''
  Phys.\ Rev.\  D {\bf 74}, 026002 (2006)
  [arXiv:hep-th/0602107];\\
C.~S.~Chu and P.~M.~Ho,
  ``Time-dependent AdS/CFT duality and null singularity,''
  JHEP {\bf 0604}, 013 (2006)
  [arXiv:hep-th/0602054].

\bibitem{Hertog}
T.~Hertog and G.~T.~Horowitz,
  ``Holographic description of AdS cosmologies,''
  JHEP {\bf 0504}, 005 (2005)
  [arXiv:hep-th/0503071];\\
T.~Hertog and G.~T.~Horowitz,
  ``Towards a big crunch dual,''
  JHEP {\bf 0407}, 073 (2004)
  [arXiv:hep-th/0406134];\\
D.~Bak,
  ``Dual of big-bang and big-crunch,''
  Phys.\ Rev.\  D {\bf 75}, 026003 (2007)
  [arXiv:hep-th/0603080].

\bibitem{Turok}
N.~Turok, M.~Perry and P.~J.~Steinhardt,
  ``M theory model of a big crunch / big bang transition,''
  Phys.\ Rev.\  D {\bf 70}, 106004 (2004)
  [Erratum-ibid.\  D {\bf 71}, 029901 (2005)]
  [arXiv:hep-th/0408083].

\bibitem{She}
J.~H.~She,
  ``A matrix model for Misner universe,''
  JHEP {\bf 0601}, 002 (2006)
  [arXiv:hep-th/0509067].

\bibitem{Joanna}
J.~L.~Karczmarek and A.~Strominger,
  ``Matrix cosmology,''
  JHEP {\bf 0404}, 055 (2004)
  [arXiv:hep-th/0309138].

\bibitem{Craps}
B.~Craps, S.~Sethi and E.~P.~Verlinde,
  ``A matrix big bang,''
  JHEP {\bf 0510}, 005 (2005)
  [arXiv:hep-th/0506180];\\
B.~Craps, A.~Rajaraman and S.~Sethi,
  ``Effective dynamics of the matrix big bang,''
  Phys.\ Rev.\  D {\bf 73}, 106005 (2006)
  [arXiv:hep-th/0601062];\\
E.~J.~Martinec, D.~Robbins and S.~Sethi,
  ``Toward the end of time,''
  JHEP {\bf 0608}, 025 (2006)
  [arXiv:hep-th/0603104];\\
T.~Ishino, H.~Kodama and N.~Ohta,
  ``Time-dependent solutions with null Killing spinor in M-theory and
  superstrings,''
  Phys.\ Lett.\  B {\bf 631}, 68 (2005)
  [arXiv:hep-th/0509173];\\
T.~Ishino and N.~Ohta,
  ``Matrix string description of cosmic singularities in a class of
  time-dependent solutions,''
  Phys.\ Lett.\  B {\bf 638}, 105 (2006)
  [arXiv:hep-th/0603215].

\bibitem{tachyoncond}
J.~McGreevy and E.~Silverstein,
  ``The tachyon at the end of the universe,''
  JHEP {\bf 0508}, 090 (2005)
  [arXiv:hep-th/0506130];\\
Y.~Nakayama, S.~J.~Rey and Y.~Sugawara,
  ``The nothing at the beginning of the universe made precise,''
  arXiv:hep-th/0606127;\\
S.~Hellerman and I.~Swanson,
  ``Cosmological solutions of supercritical string theory,''
  arXiv:hep-th/0611317.

\bibitem{tachyonrev}
M.~Headrick, S.~Minwalla and T.~Takayanagi,
  ``Closed string tachyon condensation: An overview,''
  Class.\ Quant.\ Grav.\  {\bf 21}, S1539 (2004)
  [arXiv:hep-th/0405064];\\
E.~J.~Martinec,
  ``Defects, decay, and dissipated states,''
  arXiv:hep-th/0210231.

\bibitem{Simeon}
S.~Hellerman and X.~Liu,
  ``Dynamical dimension change in supercritical string theory,''
  arXiv:hep-th/0409071;\\
S.~Hellerman and I.~Swanson,
  ``Dimension-changing exact solutions of string theory,''
  arXiv:hep-th/0612051.

\bibitem{Zwiebach}
H.~Yang and B.~Zwiebach,
  ``Rolling closed string tachyons and the big crunch,''
  JHEP {\bf 0508}, 046 (2005)
  [arXiv:hep-th/0506076];\\
H.~Yang and B.~Zwiebach,
  ``A closed string tachyon vacuum?,''
  JHEP {\bf 0509}, 054 (2005)
  [arXiv:hep-th/0506077].

\bibitem{BV} 
  R.~H.~Brandenberger and C.~Vafa, 
  ``Superstrings In The Early Universe,'' 
  Nucl.\ Phys.\ B {\bf 316}, 391 (1989). 
 
\bibitem{Perlt} 
J.~Kripfganz and H.~Perlt, 
  ``Cosmological Impact Of Winding Strings,'' 
  Class.\ Quant.\ Grav.\  {\bf 5}, 453 (1988). 

\bibitem{str} 
  T.~Battefeld and S.~Watson, 
  ``String gas cosmology,'' 
  arXiv:hep-th/0510022. 

\bibitem{rbr}
  R.~H.~Brandenberger, 
  ``Challenges for string gas cosmology,'' 
  arXiv:hep-th/0509099;\\
R.~H.~Brandenberger,
  ``Conceptual problems of inflationary cosmology and a new approach to
  cosmological structure formation,''
  arXiv:hep-th/0701111;\\
R.~H.~Brandenberger,
  ``String gas cosmology and structure formation: A brief review,''
  arXiv:hep-th/0702001.

\bibitem{Hagedorn} 
  R.~Hagedorn, 
  ``Statistical Thermodynamics Of Strong Interactions At High-Energies,'' 
  Nuovo Cim.\ Suppl.\  {\bf 3}, 147 (1965). 
 
\bibitem{Betal}
R.~H.~Brandenberger {\it et al.},
  ``More on the spectrum of perturbations in string gas cosmology,''
  JCAP {\bf 0611}, 009 (2006)
  [arXiv:hep-th/0608186].

\bibitem{TV} 
  A.~A.~Tseytlin and C.~Vafa, 
  ``Elements of string cosmology,'' 
  Nucl.\ Phys.\ B {\bf 372}, 443 (1992) 
  [arXiv:hep-th/9109048]. 

\bibitem{ns5}
A.~Sen,
  ``Strong - weak coupling duality in four-dimensional string theory,''
  Int.\ J.\ Mod.\ Phys.\  A {\bf 9}, 3707 (1994)
  [arXiv:hep-th/9402002];\\
J.~H.~Schwarz and A.~Sen,
  ``Duality symmetries of 4-D heterotic strings,''
  Phys.\ Lett.\  B {\bf 312}, 105 (1993)
  [arXiv:hep-th/9305185].


\bibitem{Ven} 
G.~Veneziano, 
  ``Scale factor duality for classical and quantum strings,'' 
  Phys.\ Lett.\ B {\bf 265}, 287 (1991). 

\bibitem{Subodh}
  S.~P.~Patil, 
  ``Moduli (dilaton, volume and shape) stabilization via massless F and D 
  string modes,'' 
  arXiv:hep-th/0504145. 

\bibitem{Cremonini}
S.~Cremonini and S.~Watson,
  ``Dilaton dynamics from production of tensionless membranes,''
  Phys.\ Rev.\  D {\bf 73}, 086007 (2006)
  [arXiv:hep-th/0601082].

\bibitem{Rador}
T.~Rador,
  ``T and S dualities and the cosmological evolution of the dilaton and the
  scale factors,''
  arXiv:hep-th/0701029.

\bibitem{Stephon}
S.~Alexander and D.~Vaid,
  ``A fine tuning free resolution of the cosmological constant problem,''
  arXiv:hep-th/0702064.

\bibitem{RHBrev5} 
R.~H.~Brandenberger,
  ``Back reaction of cosmological perturbations and the cosmological constant
  arXiv:hep-th/0210165.

\bibitem{NBV} 
  A.~Nayeri, R.~H.~Brandenberger and C.~Vafa, 
  ``Producing a scale-invariant spectrum of perturbations in a Hagedorn phase 
  of string cosmology,''
 Phys.\ Rev.\ Lett.\  {\bf 97}, 021302 (2006)   [arXiv:hep-th/0511140]. 
 
\bibitem{BNPV2}
R.~H.~Brandenberger, A.~Nayeri, S.~P.~Patil and C.~Vafa,
  ``String gas cosmology and structure formation,''
  arXiv:hep-th/0608121.

\bibitem{BNPV1} 
R.~H.~Brandenberger, A.~Nayeri, S.~P.~Patil and C.~Vafa, 
  ``Tensor modes from a primordial Hagedorn phase of string cosmology,'' 
  arXiv:hep-th/0604126. 
 
\bibitem{KKLM}
N.~Kaloper, L.~Kofman, A.~Linde and V.~Mukhanov,
  ``On the new string theory inspired mechanism of generation of  cosmological
  perturbations,''
  JCAP {\bf 0610}, 006 (2006)
  [arXiv:hep-th/0608200].

\bibitem{MFB} 
  V.~F.~Mukhanov, H.~A.~Feldman and R.~H.~Brandenberger, 
  ``Theory Of Cosmological Perturbations. Part 1. Classical Perturbations. Part 
  2. Quantum Theory Of Perturbations. Part 3. Extensions,'' 
  Phys.\ Rept.\  {\bf 215}, 203 (1992). 

\bibitem{RHBrev2} 
  R.~H.~Brandenberger, 
  ``Lectures on the theory of cosmological perturbations,'' 
  Lect.\ Notes Phys.\  {\bf 646}, 127 (2004) 
  [arXiv:hep-th/0306071]. 
 
\bibitem{Bogo}
A.~R.~Bogojevic,  
``String Inspired Cosmology: Why Space is Three-Dimensional,''
Brown University preprint BROWN-HET-691 (1988).  
 
\bibitem{Tirtho1}
T.~Biswas, A.~Mazumdar and W.~Siegel,
  ``Bouncing universes in string-inspired gravity,''
  JCAP {\bf 0603}, 009 (2006)
  [arXiv:hep-th/0508194].

\bibitem{Tirtho2}
 T.~Biswas, R.~Brandenberger, A.~Mazumdar and W.~Siegel,
  ``Non-perturbative gravity, Hagedorn bounce and CMB,''
  arXiv:hep-th/0610274.

\end{thebibliography}
\end{document}